# High Precision Ionization Chamber for Relative Intensity Monitoring of Synchrotron Radiation

S.N. Ahmed*, H.-J. Besch[†], A.H. Walenta[†], N. Pavel[†] and W. Schenk[†]
*MPI for Physics, Munich, Germany. [†]Physics Dept., Siegen University, Germany.

## Abstract

A single channel, high precision ionization chamber has been built for monitoring the relative intensity of X-rays in the energy range above 5 keV. It can be used in experiments, such as EXAFS, where simultaneous high precision monitoring of the relative intensity during the actual experiment is required. In this paper the construction of the chamber and its performance during test measurements with an X-ray tube are presented.

## 1. Introduction

In many experiments with synchrotron radiation, such as EXAFS, the intensity of incoming beam must be simultaneously monitored before the beam passes through the sample[1]. The relative accuracy of this monitoring is a critical parameter in the sense that it influences the results of experiments being performed. Such experiments can be substantially improved by having a high relative accuracy of the beam monitoring of the order of $10^{-4}$, while the attenuation of the original beam has to be sufficiently weak in order not to disturb the experiment being performed. These two conflicting requirements make the design of such a system difficult, especially for low energy photons (around 5keV) which are relatively easily absorbed. The only detection medium which could be used for low energy photons is gas because solid materials attenuate such photons to an unacceptable level. Such considerations led to the development of an ionization chamber for monitoring the relative intensity of synchrotron light. The two competing parameters mentioned above have been optimized to operate the monitor in the energy range above 5keV. The chamber has been built and tested using a 2kW X-ray tube having molybdenum anode with characteristic energy of 17.44keV. This paper presents a brief description of the detector design, construction, and analysis of test measurements.

## 2. Chamber Design

The mechanical design of this single channel ionization chamber is shown in fig.1. The sensitive volume of the chamber consists of two 8mm thick regions separated by a central readout electrode having 7mm diameter and bounded by two field forming electrodes. In order to smoothen the field at the edges of circular electrodes, two field rings have been provided in each region. These rings are kept at half of the potential of the field forming electrodes. The whole setup is kept in an aluminum cylindrical vessel with end gaps of 8mm on each side. All the electrodes and windows are made of $6\mu m$ thick mylar foil metallized with $0.3\mu m$ thick aluminum. The foils have been stretched and glued on aluminum rings which are then mounted to one of the lids of the vessel with screws. A steady gas flow in the chamber is ensured through holes in the ring. A metalized layer of aluminum around the central electrode serves as guard ring. The electrical connections to the electrodes have been made using glue and conductive silver paint. A printed circuit board has been provided to facilitate connections with readout electronics and HV supply.

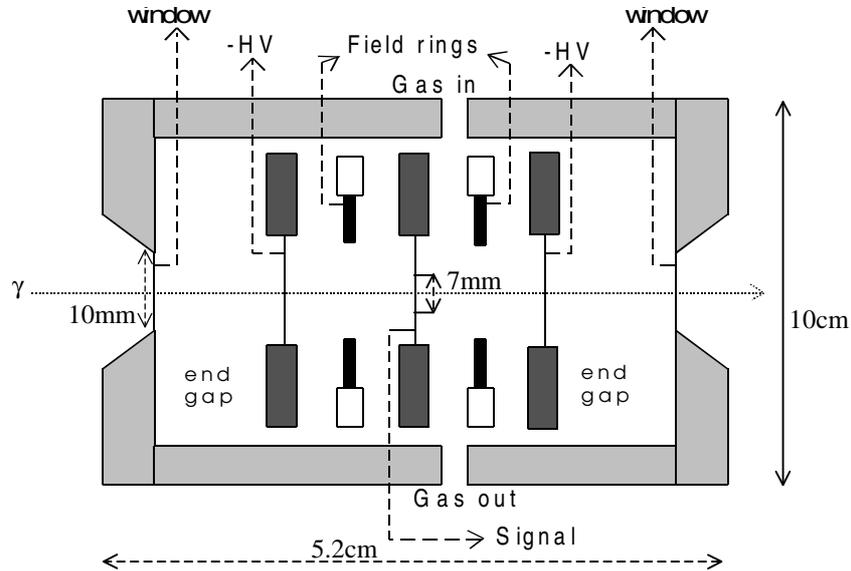

Fig.1: Schematic of beam monitor (expanded in horizontal direction to show details). All the four regions are equally spaced (8mm) in horizontal direction. The field rings are kept exactly in the middle of central and outer electrodes.

The distances between the electrodes were kept smaller than the diameter of the electrodes. Together with the field rings between electrodes, this ensures parallel field lines in the active parts of the chamber. This is necessary to collect efficiently all the electron-ion pairs generated in the active regions. The HV electrodes are placed in a position of equilibrium of electrostatic force in order to avoid voltage dependent deformations and consequently inaccuracies of the measurement.

The vessel is sealed by O-rings. The gas inlet and outlet have been provided on top of the vessel. The chamber operates at atmospheric pressure mainly because of the very thin windows which can withstand only very low pressure gradients.

3. Choice of gas

The filling gas in the chamber should be capable of generating a sufficiently large number of electron-ion pairs in order to ensure good signal to noise ratio. At the same time the beam should not be attenuated to an undesirable level. In general an absorption of 5-20% of the original beam is considered acceptable. For a particular experiment, the choice of gas depends on the required resolution and acceptable attenuation of beam. For this work the chamber was tested with gas mixtures of 95%Ar+5%$CO_2$ and 90%Ar+10%$CH_4$.

4. Electronics

The readout electronics chain (see fig.2) consists of a high precision integrating amplifier IVC102[2], a 16 bit 100kHz analog to digital converter ADS7805[3] and a personal computer with suitable interface card to read the digitized data. To provide the necessary switching digital signals to drive integrator and ADC, a universal controller was used.

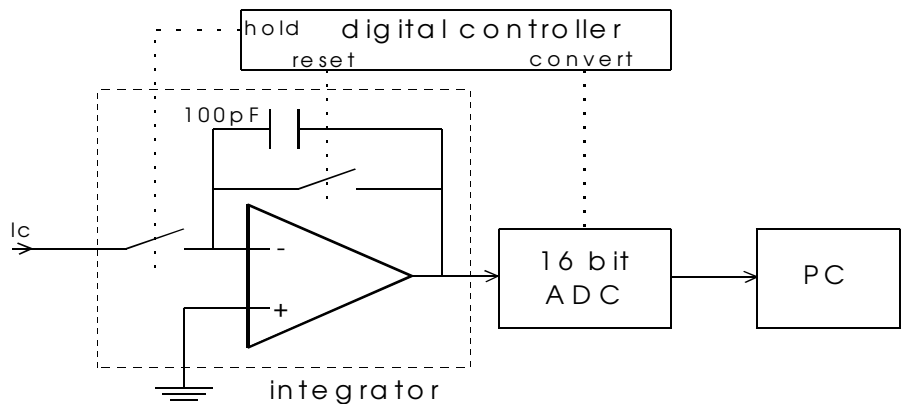

Fig.2: Readout electronics

The controller operates with a 1MHz crystal oscillator and is programmable. For test measurements the integration time was set to 1mS. The noise of the integrator[2] as given by the supplier is 40μV for an integrator capacitance of 100pF and the digitization error of the ADC[3] is 26.4μV. This corresponds to an error equivalent of 35.9 absorbed photons for 1mS integration time. Considering Poisson statistics this is adequate for a precision of about 0.1% or better.

5. Experimental setup for test measurements

The experimental setup is shown in fig.3. A 2kW crystallographic X-ray tube with a molybdenum anode, which has a characteristic energy of 17.44keV, has been used for the test measurements. The beam was collimated with a very narrow lead collimator. The relative intensity of X-rays could be controlled by tube current because from previous experience[4] it was known that, for the particular tube used, the intensity is directly proportional to the tube current. Two different gas mixtures, 95%Ar+5%$CO_2$ and 90%Ar+10%$CH_4$, at atmospheric pressure were used for the test measurements.

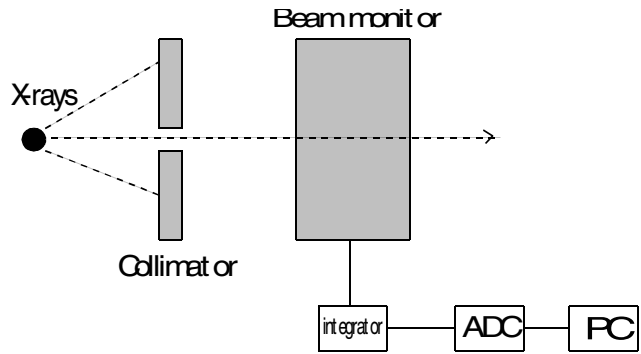

Fig.3: Experimental setup

## 6. Results

### 6.1. Ionization Chamber Plateau

Fig.4 shows the ionization chamber plateaus for different X-ray intensities. Careful designing, positioning and mounting of electrodes of chamber result in very flat chamber plateaus at different photon fluxes provided the right gas mixture is chosen. During present work the chamber was operated with two different gas mixtures. A mixture of 90%Ar+10%$CH_4$ resulted in plateau slopes of less than 2% per 1000 volts while a mixture of 95%Ar+5%$CO_2$ produced slopes of roughly 3% per 1000 volts. This means that the voltage the voltage has to be kept constant at level of a few Volts to ensure a stability of $10^{-4}$.

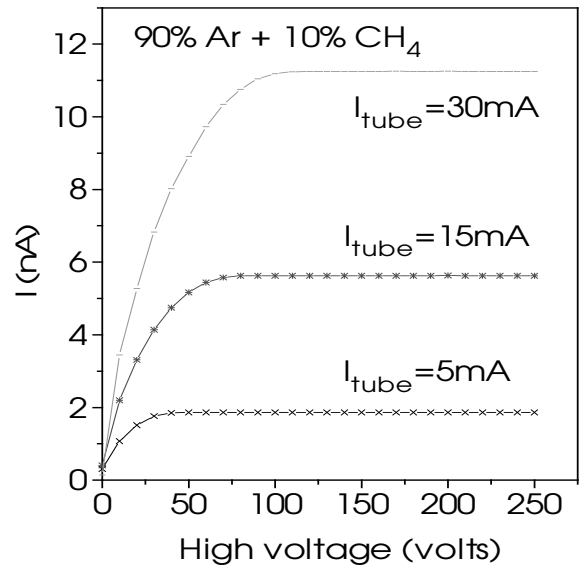

Fig.4: Ionization chamber plateaus at different X-ray intensities

*6.2. Linearity*

For a proper performance of the monitor the detector signal must depend linearly on the incident photon flux. To check the linearity of the detector response, the digitized voltage signal from the detector was measured as a function of the X-ray tube current.

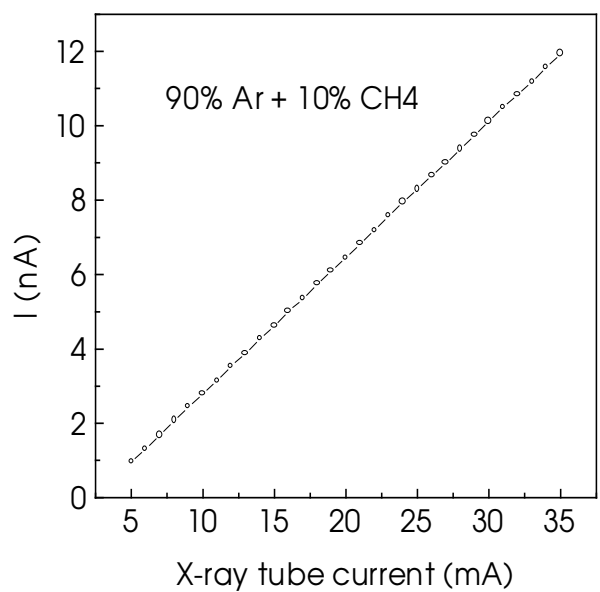

Fig.5: Test of the linearity of the chamber; ionization current as a function of the X-ray tube current.

At each step of the X-ray tube current, 1024 voltage measurements were performed. The average of these was then converted into equivalent average ionization current $<I_{ion}>$ using the known conversion factors of ADC and Integrator. For all measurements the chamber was biased at 200 V using a stable high voltage supply. Fig.5 shows the plot of $<I_{ion}>$ versus the current of X-ray tube. A quadratic fit to the experimental values gave a maximum quadratic contribution of 1.2% of the linear contribution. Keeping in view the small non-linearity in the X-ray tube itself, this is negligibly small and therefore the system can be said to behave linearly with respect to incident neutron flux.

*6.3. $\sigma^2$-N dependence*

A series of measurements were performed to check the noise behavior of the whole detection system with respect to number of absorbed photons. If the precision of the measurement is determined mainly by the Poisson statistics of the absorbed photons then, for a approximately monochromatic beam of incident photons, one would expect a linear dependence of the variance of measurements with respect to average number of absorbed photons, i.e.,

$$\sigma_{abs}^2 = E_\gamma^2 \cdot N_{abs}$$

Here $N_{abs}$ is the number of absorbed photons of energy $E$ and $\sigma_{abs}$ is the standard deviation of the absorbed energy. This simple relation holds for the analysis of the ionization current measurements because the number of ion pairs per photon is large and their fluctuations are further reduced by Fano factor.

The X-ray tube used for the tests has, however, is not monochromatic. Thus one has to replace $E_\gamma^2$ by $<E_\gamma^2>$.

From the spread of the X-ray spectrum, $\sigma^2 = <E^2> - <E>^2$, one obtains for the measured variance of the absorbed energy

$$\sigma_{abs}^2 = <E^2> N_{abs} = <E>^2 N_{abs} (1 + \sigma^2/<E>^2)$$

Spread and mean value have been determined from the measurement of the energy spectrum of the X-ray source.

In addition to the Poisson fluctuations, there are other additive noise sources in the system, e.g., electronics noise and digitization error. Since these noise sources are statistically independent, the total variance of the measured counts is given by adding the corresponding variances in quadrature.

$$\sigma_m^2 = \sigma_{add}^2 + <E_\gamma>^2 \cdot N_{abs} \cdot \left[1 + \frac{\sigma_E^2}{<E_\gamma>^2}\right]$$

Thus plotting $\sigma_m^2$ versus $N_{abs}$ should give a straight line with the y-intercept being the contribution of other noise sources in the system. From the slope the effective energy of absorbed photons can be determined. The slope of the above equation depends also on the spread in the spectrum of absorbed photons. The spectrum of incident photons was determined by a proportional counter and from that the spectrum of absorbed photons was calculated using the mass absorption coefficients of the gas in the beam monitor. The spread in the spectrum of absorbed photons was found to be 3.65keV. This was used to determine the effective energy of absorbed photons which turned out to be 12.46keV. This value which is somewhat lower than the average energy of incident photons (15 keV) can be understood from the stronger absorption of lower energy photons in the active volume of the chamber as compared to higher energy photons. The average energy of absorbed photons was also calculated from the measured spectrum of absorbed photon. It was found to be 12.25keV giving a relative difference of about 2% as compared to the measured value.

Fig.6 shows the variance of the absorbed photons ($\sigma^2$) as a function of the average number of absorbed photons calculated using the above two relations. A contribution of a 100 Hz ripple on the signal, which was caused by power supply of the X-ray tube, was filtered out during the offline data reduction by Fourier filtering in order not to falsify the results. It should be noted that the values represent the number of photons absorbed in 1mS integration time.

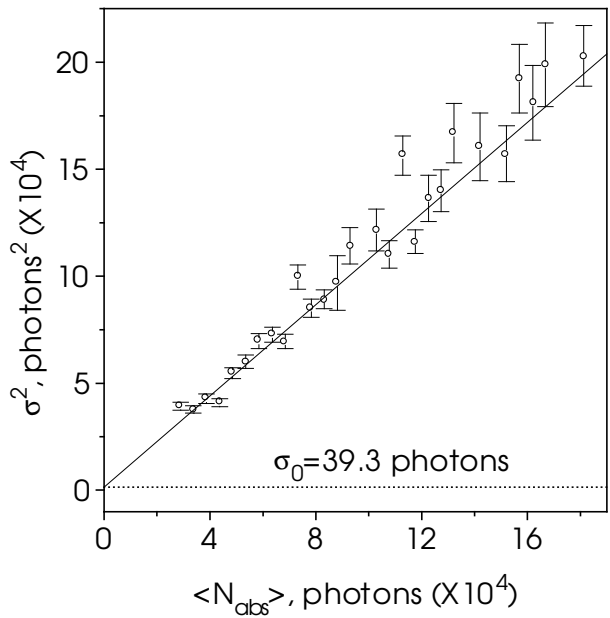

Fig.6: Dependence of variance of measurements on the average number of absorbed photons and a straight line fit to the data.

Experimentally the variance $\sigma^2_{Nc}$ of the measured voltage in ADC units is determined from the variance of 1024 consecutive measurements under the same conditions. The value of $\sigma^2$ of $N_{abs}$ is then given by:

$$\sigma^2_{N_{\gamma,abs}} = \sigma^2_{N_c} \cdot \left[ \frac{W \cdot b \cdot C}{e \cdot \langle E_{abs} \rangle \cdot \tau} \right]^2$$

where W is the average energy for the production of en electron-ion pair, e is the electron charge in Coulomb, b is the conversion factor for ADC units into Volts, C is the integration capacity and $\tau$ is the integration time.

The additional noise of the system as calculated previously from the y- intercept of the straight line fit is equivalent to 38.4±21.1 photons being well compatible within the statistical error with the expected electronics noise equivalent of 35.9 absorbed photons.

In most cases the beam intensity is monitored in intervals of the order of a second, thus taking the average $<N_{abs}>$ of 1000 single measurements. The relative error of this average $<N_{abs}>$ is shown as a function of $N_{abs}$ in fig. 7, together with the relative error expected from Poisson fluctuations only (solid line). This demonstrates again that the beam monitor operates at the limit of quantum fluctuations and that a relative accuracy of the order of $10^{-4}$ can be achieved. At low beam intensities averaging over more than a second would be necessary for the same accuracy, while at beam intensities larger than about $10^{11}$ photon/sec measurements of 1ms are sufficient.

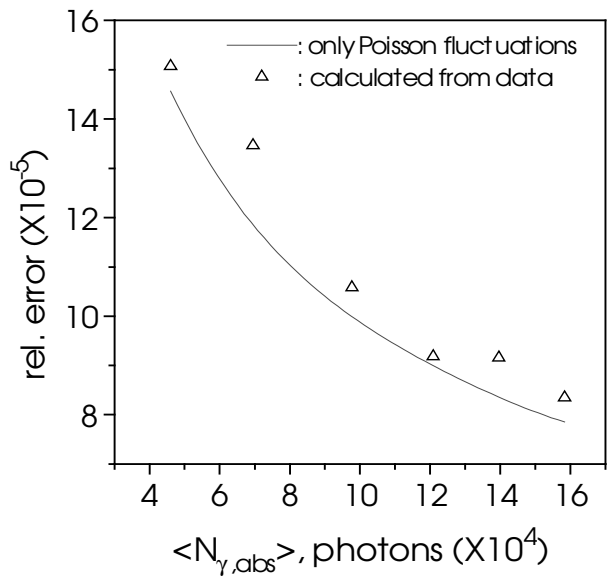

Fig.7: Relative errors in mean number of absorbed photons.

6.4. Systematic uncertainties

The use of this beam monitor at a synchrotron light source might give rise to some systematic uncertainties because of the available higher flux of incoming photons. These include the effects due to accumulation of space charge and change of the internal or external pressure. The effect of space charge for this particular geometry has been analytically worked out (see ref.[6]) and the result is shown in fig.8. It can be seen that up to $10^{10}$ photons/sec the space charge does not deteriorate the electric field intensity inside the active volume considerably, i.e., it remains uniform to a good extent and the charge collection efficiency of the monitor is not degraded. However above this value, the effect becomes sizable and some software filter scheme would be needed to make reliable analysis of the measurements.

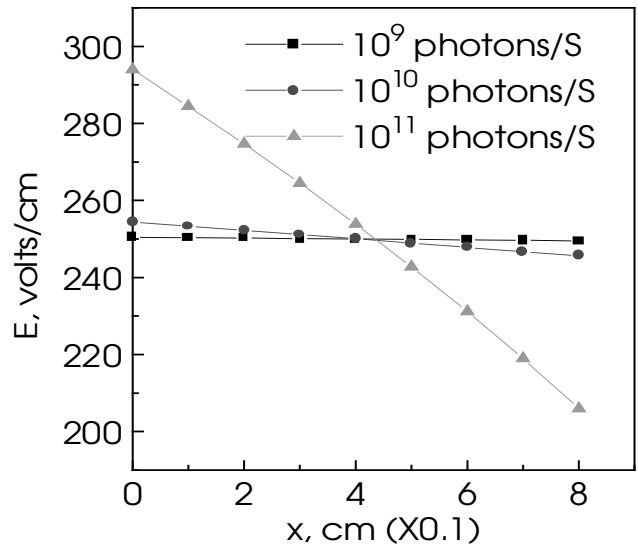

Fig.8. Distortion of electric field intensity in the active volume of the beam monitor due to accumulation of space charge (ref.[6]).

The change of pressure inside or outside the beam monitor would also affect its performance. These barometric shifts can arise from change in atmospheric temperature and microphony in the environment of the experiment or from the gas flow inside the chamber. In ref.[6], analytic calculations have been performed to estimate the magnitude of uncertainty due to such barometric shifts. It turned out that at small pressure changes, the effect is not very significant: if the monitor is being operated at 1 bar, filled with 100% Argon and exposed to a flux of $10^{10}$ photons/sec, then a change of pressure of 2 mbar would appear as a relative change of $\sim 1 \times 10^{-3}$ in measured ionization current. However if the pressure changes by as much as 10mbar then the error in measured ionization current is about an order of a magnitude higher. Therefore care would be necessary to keep the gas flow rate constant and monitor the atmospheric conditions during the experiment as well.

7. Conclusions

A high precision X-ray beam monitor for applications in synchrotron radiation experiments was built and the results of performance tests at a X ray tube have been presented.

The beam monitor is a single channel ionization chamber working at normal gas pressure. It is designed to operate in the energy range of 5-20 keV without a significant attenuation of the original beam. The linearity of the ionization chamber signal was shown and the slope of the plateau was found to be less than 2% per 1000 V.

The tests have demonstrated that this beam monitor operates at the limit of quantum fluctuations. During the tests a relative accuracy of $3\times10^{-4}$ for the Intensity measurement could be achieved. Extrapolating this result to higher beam intensities produced in synchrotrons this implies that in this application a relative precision of the beam monitoring of $10^{-4}$ is well achievable.

Space charge effects should not deteriorate the accuracy to a level of more than $10^{-4}$ when exposed to a photon flux up to the order of $10^{10}$ photons/sec. When using the beam monitor in experimental environment much care has to be taken to avoid uncontrolled variations of the inner and outer pressure.

Acknowledgements

We thank the mechanical workshop of Siegen University for building the mechanics of the chamber timely and carefully. Special thanks are due to Mr. G. Iksal for his valuable support for designing and realizing a part of the electronics chain of the system.